\documentclass[aps,pra,twocolumn,superscriptaddress,groupedaddress]{revtex4}
\usepackage{graphicx}  
\usepackage{dcolumn}   
\usepackage{bm}        
\usepackage{amssymb}   

\begin{document}

\title{Multiphoton Correlations between Quantum Images}

\author{Serge Massar}
\affiliation{Laboratoire d'Information Quantique CP224, Universit{\'e} libre de Bruxelles (ULB), Av. F. D. Roosevelt 50, B-1050 Bruxelles, Belgium}
\author{Fabrice Devaux}
\author{Eric Lantz}
\affiliation{Institut FEMTO-ST,
D{\' e}partement d’Optique, UMR 6174 CNRS,
Universit{\' e} de Franche-Comt{\' e}, 15 B rue des Montboucons, 25030 Besan{\c c}on, France}
\date{\today}

\begin{abstract}
Experimental demonstrations of entangled quantum images produced through parametric downconversion have so far been confined to studying two photon correlations. Here we show that multiphoton correlations between quantum images are accessible experimentally and exhibit many new features including being sensitive to the phase of the 
bi-photon wavefunction. As a concrete example, we consider a modification of existing quantum imaging experiments  in which the CCD cameras  are moved out of focus,  provide detailed analytical predictions for the resulting 4 photon intereferences, and support these by numerical simulations.
The proposed experiment can also be interpreted as entanglement swapping: Bob's photons are initially unentangled, but the joint detection of Alice's photons projects Bob's photons onto an entangled state.
The general approach proposed here  can  be extended to other quantum optics experiments involving high dimensional entanglement. 
\end{abstract}

\maketitle

\section{Introduction}

Since the seminal works of Clauser \cite{Clauser72} and Aspect \cite{Aspect82}, entangled photons have been one of the workhorses of quantum information sciences. Nowadays high dimensional entangled photon pairs can be routinely produced in the laboratory, using  different degrees of freedom such as angular momentum \cite{Dada11, Krenn14}, time-energy \cite{Olislager10,Xie15,Imany18,Chang21}, position-momentum \cite{Howell94,Edgar12,Moreau14,Devaux19}, path entanglement (using integrated optics) \cite{Wang18}, or multiple degrees of freedom simultaneously \cite{Barreiro05}. The number of modes that can be entangled can reach hundreds, or even thousands, see   e.g.\cite{Edgar12,Moreau14,Krenn14,Devaux19,Chang21}. These experiments have focused on the correlations between two entangled photons. Here we show that if one extends them to the study of multiphoton correlations, then novel phenomenology and  interference patterns emerge. These new features are experimentally accessible with current technology, as they already appear in 4 photon correlations. In the same way that the quantum teleportation experiment of Bouwmeester et al. \cite{Zeilinger97} revolutionized quantum optics, we expect the present proposal to considerably broaden the scope and  interest of high dimensional photonic entanglement.

An important inspiration for the present work is boson sampling \cite{Aaronson11}, see the experimental realizations of \cite{Broome13,Spring13,Tillmann13,Paesani19,Zhong20}. On the one hand boson sampling provides the theoretical framework for describing  multiphoton correlations. On the other hand the computational complexity arguments of \cite{Aaronson11} show that as the number of modes and the number of photons increases, the  correlation pattern  become exceedingly complex and impossible to simulate efficiently on a classical computer. But for moderate number of photons (say 4 or 6), while this complexity already shows up, it should be possible to fully investigate the system experimentally. 
 The present work is most closely related to the extension of boson sampling to gaussian bi-partite states \cite{Chakhmakhchyan17,Grier21}, and to the low optical depth boson sampling of \cite{Raoul21}.

\begin{figure}
\begin{center}
  \includegraphics[width=1.0\linewidth]{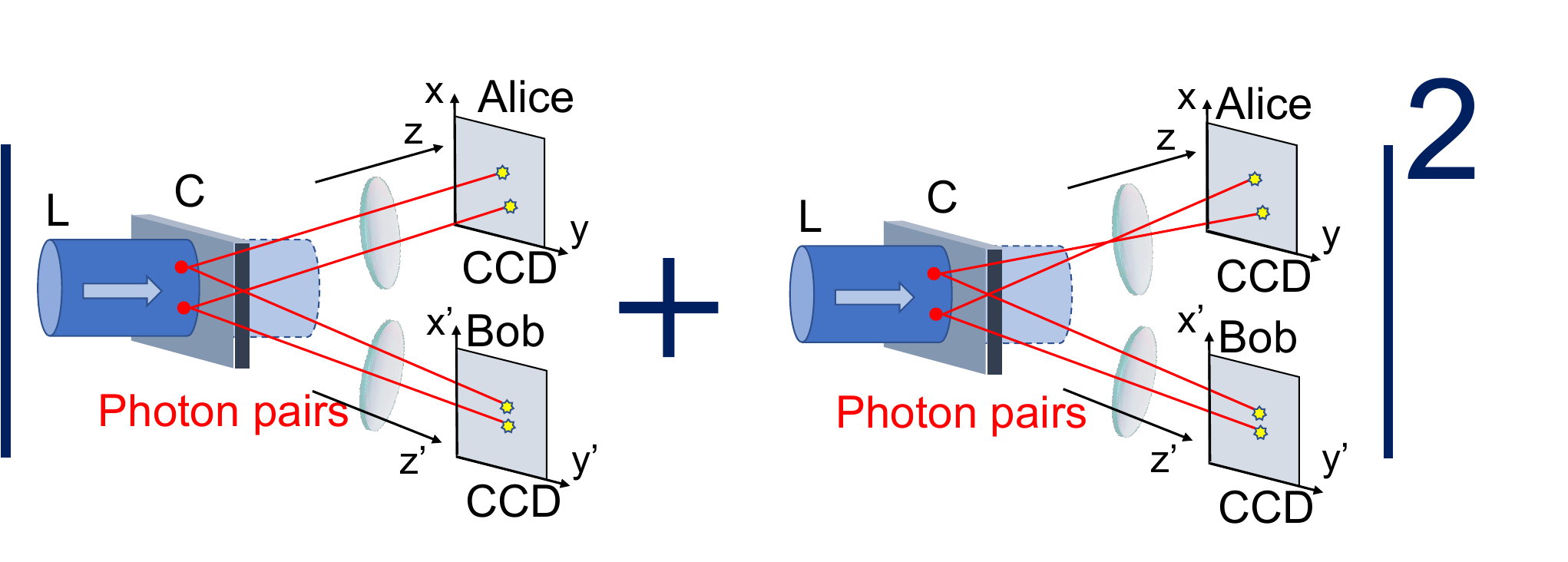}
\end{center}
  \caption{Schematic of the proposed setup and interferences between 4 photon correlations. A spatially extended, pulsed, pump laser (L)  generates photon pairs in a nonlinear crystal (C) 
  through SPDC. The signal and idler photons (red lines) are imaged onto Alice's and Bob's CCD cameras. 
  If two photon are detected on Alice's camera and two photons are detected on Bob's camera, then we don't know which photon is the partner of which photons. The amplitudes for the two processes interfere, and must be summed when computing the probability of this event, see Eq.
   (\ref{Eq:4photon}). This is  indicated schematically in the figure by 
  the ``$+$''
 sign between the two amplitudes which must be taken before squaring. $x,y,z$ and $x',y',z'$ are the coordinates used in the main text.   The lens in the image indicate schematically that an imaging system system is used, specifically either a 2-f or 4-f optical system, slightly defocussed as measured by $z$ and $z'$. 
}
  \label{fig:setup}
\end{figure}

For definitness we  illustrate our approach in the case of spatially entangled photons, as realised in \cite{Edgar12,Moreau14}, and schematized in Fig. \ref{fig:setup}. A spatially extended, pulsed, pump laser illuminates a thin nonlinear crystal in which photon pairs are produced by Spontaneous Parametric Down Conversion (SPDC) using Type II phase matching. The signal and idler photons are not colinear and are imaged separately on Alice and Bob's cameras. Single photon resolution on each pixel of the camera is achieved by using electron multiplying charge-coupled
devices (EMCCD). Such quantum imaging experiments were introduced theoretically in \cite{Kolobov89}. Using CCD cameras, they have been applied to
demonstrations of the Einstein-Podolsky-Rosen
(EPR) paradox\cite{Moreau12,Moreau14}, ghost imaging\cite{Morris15},
quantum adaptive optics\cite{Defienne18}, quantum holography\cite{Devaux19},
sub-shot-noise imaging \cite{Brida10,Toninelli17}, quantum imaging with
undetected photons \cite{Lemos14}, see \cite{Moreau19} for a review. Note that additional optical elements, such as a Spatial Light Modulator (SLM), or a diffuser \cite{Defienne18,SLMD21}, can be inserted between the source and the CCD if desired. Ghost imaging with entanglement-swapped photons was reported in \cite{Bornman19}
demonstrating the feasibility of multiphoton quantum imaging experiments.

Here we consider multiphoton correlations on the image planes. If the produced photons are indistinguishable (except for the position-momentum degree of freedom), then we don't know which photon detected on Alice's camera is the partner of which photon detected on Bob's camera. The probability for a specific detection event is obtained by summing all the possible pairings of signal and idler photons, as illustrated in Fig. \ref{fig:setup} in the case of 4 photon correlations.


\section{Multiphoton correlations in quantum images}

We treat the pump beam classically which implies that the quantum state of the signal and idler photons is gaussian and can be written as 
\begin{equation}
\vert \Psi \rangle = N \exp \left( \int\int d{\bf  x}  d {\bf  x'}  \Phi({\bf  x},{\bf  x'})
a_{\bf  x}^\dagger a_{\bf  x'}^\dagger \right) \vert 0 \rangle\ ,
\label{Eq:GaussianPsi}
\end{equation}
where $\Phi({\bf  x},{\bf  x'})$ is the biphoton wavefunction, with ${\bf  x} = (x,y)$ and ${\bf  x'} = (x',y')$ the positions on Alice and Bob's image planes (we denote throughout  Alice's (Bob's) variables with unprimed (primed) letters),  $a_{\bf  x}^\dagger$ and $a_{\bf  x'}^\dagger $ are creation operators for photons at  ${\bf  x}$ and ${\bf  x'}$,  $\vert 0 \rangle$ is the vacuum state, and $N$ is a normalisation factor.

The probability of detecting $n$ photons on Alice's camera at positions 
${\bf X}_{(n)}={\bf  x}_1,\cdots,{\bf  x}_n$ and $n$ photons on Bob's camera at positions 
${\bf X}^{'}_{(n)} = {\bf  x}'_1,\cdots,{\bf  x}'_n$ is given by \cite{Chakhmakhchyan17,Grier21}
\begin{eqnarray}
P^{(2n)}({\bf X}_{(n)}; {\bf X}^{'}_{(n)} )
&=&
\vert \langle 0 \vert a_{{\bf  x}_1} \cdots a_{{\bf  x}_n} a_{ {\bf  x}'_1}\cdots 
a_{{\bf  x}'_n} \vert \Psi \rangle \vert^2
\nonumber\\
&=& \vert N \vert^2 \vert {\mathrm{Perm}}\left(
\Phi_{{\bf  x}_1,\cdots,{\bf  x}_n; {\bf  x}'_1,\cdots,{\bf  x}'_n} \right) \vert^2
\quad \  
\label{Eq:Perm}
\end{eqnarray}
where $\Phi_{{\bf  x}_1,\cdots,{\bf  x}_n; {\bf  x}'_1,\cdots,{\bf  x}'_n}$ is the $n \times n$ matrix whose $(i,j)$th entry is given by the biphoton wavefunction at positions  $({\bf  x}_i,  {\bf  x'}_j)$, i.e. by $\Phi({\bf  x}_i, {\bf  x'}_j)$, and ${\mathrm{Perm}}$ is the permanent of the matrix. 

In order to simplify expressions, we make the following approximations. First, in order to get the response of the camera we need to integrate Eq. (\ref{Eq:Perm}) over the area of each pixel. We assume that $\Phi$ varies little over the area of a pixel and therefore  omit this integration. Second we assume that  the mean number of photons is much smaller than the  number of pixels, and consequently the probability of two photons reaching the same pixel is small, and we do not consider these events.  Third we assume that losses are negligible.

Thus the probability of detecting a single pair at positions $({\bf  x}; {\bf  x'})$ is given by
\begin{equation}
P^{(2)}({\bf  x}; {\bf  x'})= \vert N \vert^2  \vert  \Phi({\bf  x}, {\bf  x'})  \vert^2\ ,
\end{equation}
and depends only on the norm of the biphoton wavefunction. But if two pairs are detected at $({\bf  x}_1,{\bf  x}_2)$ and $({\bf  x}'_1,{\bf  x}'_2)$, then the corresponding probability is given by
\begin{eqnarray}
P^{(4)}({\bf  x}_1,{\bf  x}_2; {\bf  x}'_1,{\bf  x}'_2)&=& \vert N \vert^2  \vert  \Phi({\bf  x}_1, {\bf  x}'_1) \Phi({\bf  x}_2, {\bf  x}'_2) 
\nonumber\\
& &\quad 
+ \Phi({\bf  x}_1, {\bf  x}'_2) \Phi({\bf  x}_2, {\bf  x}'_1) 
 \vert^2\ .
 \label{Eq:4photon}
\end{eqnarray}
A new interference effect arises because we do not know whether the photon detected at ${\bf  x}_1$ is the partner of the photon detected at ${\bf  x}'_1$ or at ${\bf  x}'_2$, and we must sum the amplitudes for these two processes as illustrated in Fig. \ref{fig:setup}. Therefore Eq. (\ref{Eq:4photon}) is sensitive to the phase of the biphoton wavefunction. 

\section{Defocusing the quantum images}

For the purpose of analytical predictions, we assume that the biphoton wavefunction is gaussian, which is a widely used and well justified approximation \cite{LE04,Fedorov07,CTE07}. 
At the surface of the nonlinear crystal (i.e. in the near field)   the biphoton wavefunction is thus given by
\begin{equation}
\Phi({\bf  x}, {\bf  x'}) 
\propto \exp\left( - \frac{1}{4 w_0^2}  \vert {\bf  x} + {\bf  x'}\vert^2-\frac{b^2}{4} \vert{\bf  x} - {\bf  x'}\vert^2\right) \ .
\label{Eq:Phixx'}
\end{equation}
For simplicity of notation we omit, here and in the following expressions, the constant that multiplies the exponentials in $\Phi$, and denote this by the symbol $\propto$. 
In Eq. (\ref{Eq:Phixx'}), $w_0$ is the width of the pump beam, while $b$ takes into account that the phase matching conditions are only partially enforced due to the finite thickness of the nonlinear crystal. The photon pairs are produced approximately in the same location, up to an uncertainty $1/b$. The Schmidt number of this biphoton wavefunction is $K= \frac{1}{4} \left( b w_0 + \frac{1}{b w_0} \right)^2$\cite{LE04}. We are interested in the situation where $w_0 \gg 1/b$, corresponding to a large area of illumination of the crystal and a high Schmidt number. Experimentally   the total number of position/momentum modes in entangled images of order $2000$ are reported \cite{Edgar12}, although direct measurements of the Schmidt number have yielded a lower value of order 
 $200$ \cite{Moreau14}.

The Fourier transform of Eq. (\ref{Eq:Phixx'}) gives the biphoton wavefunction in the far field
\begin{equation}
\tilde \Phi({\bf  p}, {\bf  p'}) 
\propto \exp\left( 
- \frac{ w_0^2}{4}\vert {\bf  p} + {\bf  p'}\vert^2
-\frac{1}{ 4 b^2} \vert{\bf  p} - {\bf  p'}\vert^2
\right) 
\label{Eq:Phipp'}
\end{equation}
where ${\bf  p}$   and  ${\bf  p'}$ are the transverse momenta of Alice and Bob's photons.


For interesting interferences to arise in the 4 photon coincidences Eq. (\ref{Eq:4photon}) we need a complex, oscillating biphoton wavefunction. This is not the case for 
the near and far field biphoton wavefunctions Eqs. (\ref{Eq:Phixx'}, \ref{Eq:Phipp'}) which are real and positive. But in \cite{CTE07} it was shown that as the photons propagate, the entanglement between Alice's and Bob's photons becomes encoded in the phase of the biphoton wavefunction.  This situation is  readily accessible regime experimentally: one simply needs to move the cameras out of focus. 
Note that because of the symmetry between Eqs. (\ref{Eq:Phixx'}) and (\ref{Eq:Phipp'})
one could either defocus the near field image or the far field image. Below we consider the case of defocusing the near field image.

If Alice and Bob's photons travel a distance $z$ and $z'$ from the crystal surface, then in the paraxial approximation, the biphoton wavefunction in momentum space is given by
\begin{eqnarray}
\tilde \Phi({\bf  p} , {\bf  p'};z,z') 
&\propto& \exp\left( 
- \frac{ w_0^2}{4}\vert {\bf  p} + {\bf  p'}\vert^2
-\frac{1}{ 4 b^2} \vert{\bf  p} - {\bf  p'}\vert^2
\right.
\nonumber\\
& & \quad \left.
- i \frac{z}{2k}\vert {\bf  p} \vert^2 - i \frac{z'}{2k}\vert {\bf  p'} \vert^2 
\right) 
\label{Eq:Phipp'zz'}
\end{eqnarray}
where $k$ is the longitudinal momenta of the idler and signal photons (assumed equal).
In order to simplify the 
expression for the Fourier transform of Eq. (\ref{Eq:Phipp'zz'}), we take the limit $w_0 \to \infty$ in the resulting expression, whereupon  $\Phi({\bf  x} ,{\bf  x'};z,z') $ only depends on ${\bf  x}- {\bf  x'}$, i.e. we are in the translation invariant limit. We then have
\begin{eqnarray}
\Phi({\bf  x}, {\bf  x'};z,z') 
&\propto& \exp\left( -\frac{\alpha - i \beta}{4} \vert {\bf  x} - {\bf  x'}\vert^2
\right) 
\label{Eq:Phixx'zz'alphabeta}
\end{eqnarray}
where
$\alpha = \frac{b^2}{1+Z^2}$ and
$\beta = \frac{b^2 Z }{1+Z^2}$ are real and positive with $Z= \frac{b^2 (z+z')}{2 k}$.
The uncertainty in the joint positions is  of size $1/\sqrt{\alpha}$, and increases when the defocusing (i.e. $z$ and $z'$) increase. 
 The defocusing is important when $Z\gg 1$, whereupon $\beta \gg \alpha$, and the biphoton wavefunction exhibits many oscillations within a defocusing spot.

Going back to Fig. \ref{fig:setup}, Eqs. (\ref{Eq:Phixx'},  \ref{Eq:Phipp'}, \ref{Eq:Phixx'zz'alphabeta})  correspond to the bi-photon wavefunction on the camera planes when imaging the near field (the crystal surface), the far field, and the defocused near field respectively.
The coordinates $(x,y)$ and $(x',y')$ in Fig. \ref{fig:setup} correspond to $\bf x$ and ${\bf x'}$ in the case of Eqs. (\ref{Eq:Phixx'}, \ref{Eq:Phixx'zz'alphabeta}) ) and to 
$\bf p$ and ${\bf p'}$ in the case of Eq. (\ref{Eq:Phipp'}), while the $z$ and $z'$ coordinates correspond to the degree of defocusing.

In order to obtain predictions for the 4 photon correlations, we  insert Eq. (\ref{Eq:Phixx'zz'alphabeta}) into Eq. (\ref{Eq:4photon}). One finds that the 4 photon correlation probability takes the  simple form
\begin{equation}
P^{(4)}({\bf  X}; {\bf  X'}; z,z') \propto 
\exp(- \frac{\alpha}{4}D)\left( \cosh( \frac{\alpha}{2} S) + \cos(\frac{\beta}{2} S)\right)\label{Eq:P4}
\end{equation}
where
$D({\bf  X}; {\bf  X'}) = \vert {\bf x_1} - {\bf  x_1'} \vert^2
+\vert {\bf  x_1} - {\bf  x_2'} \vert^2
+\vert {\bf  x_2} - {\bf  x_1'} \vert^2
+\vert {\bf  x_2} - {\bf  x_2'} \vert^2
$
and
$S({\bf  X}; {\bf  X'})  =
 ({\bf  x_1} - {\bf  x_2}). ({\bf  x'_1} - {\bf  x'_2})$.
When the defocusing is significant ($\beta \gg \alpha$) then the 4-photon coincidence probabilities  have strong oscillations given by the term $ \cos(\frac{\beta}{2} S)$.
In Appendix \ref{AppendixRobustness}  we 
show that the oscillating term in Eq. (\ref{Eq:P4}) (the term in $\cos(\frac{\beta}{2} S)$) is a robust prediction that does not depend on the gaussian approximation Eq. (\ref{Eq:Phixx'}). And in Appendix \ref{App:HighOrder} we  
generalize Eq. (\ref{Eq:P4}) to higher order correlations and show that the expressions for $P^{(2n)}$ are much more complex as soon as $n>2$.

\section{Numerical simulations}

In order to confirm these analytical predictions, we carried out numerical simulations using the method introduced in \cite{Brambilla04} and since used extensively, see e.g. \cite{Lantz21}. 
The idea of the simulations is to take as input for the signal and idler fields gaussian white noise with intensity corresponding to half a photon per mode. This field is numerically propagated through the system, including the non linear crystal. The obtained fields are used to obtain, after averaging over repetitions of the simulation and appropriate subtractions, expectation values of the 4 point intensity correlations
\begin{equation}
I({\bf  x}_1,{\bf  x}_2; {\bf  x'}_1,{\bf  x'}_2 )=
\langle \Psi \vert n_{{\bf  x}_1} n_{{\bf  x}_2} n_{ {\bf  x'}_1} n_{{\bf  x'}_2} \vert \Psi \rangle
\label{Eq:I-4}
\end{equation}
where $n_{x}=a^\dagger_x a_x$ is the number operator at position $x$. 
We then subtract the correlations of lower order: accidental coincidences between non twin photons and between non twin signal and idler pairs (two bunched photons in an image, that do not correspond to twin photons in the other image) to obtain the genuine 4 point intensity correlations.

\begin{figure}
\begin{center}
  \includegraphics[width=1.0\linewidth]{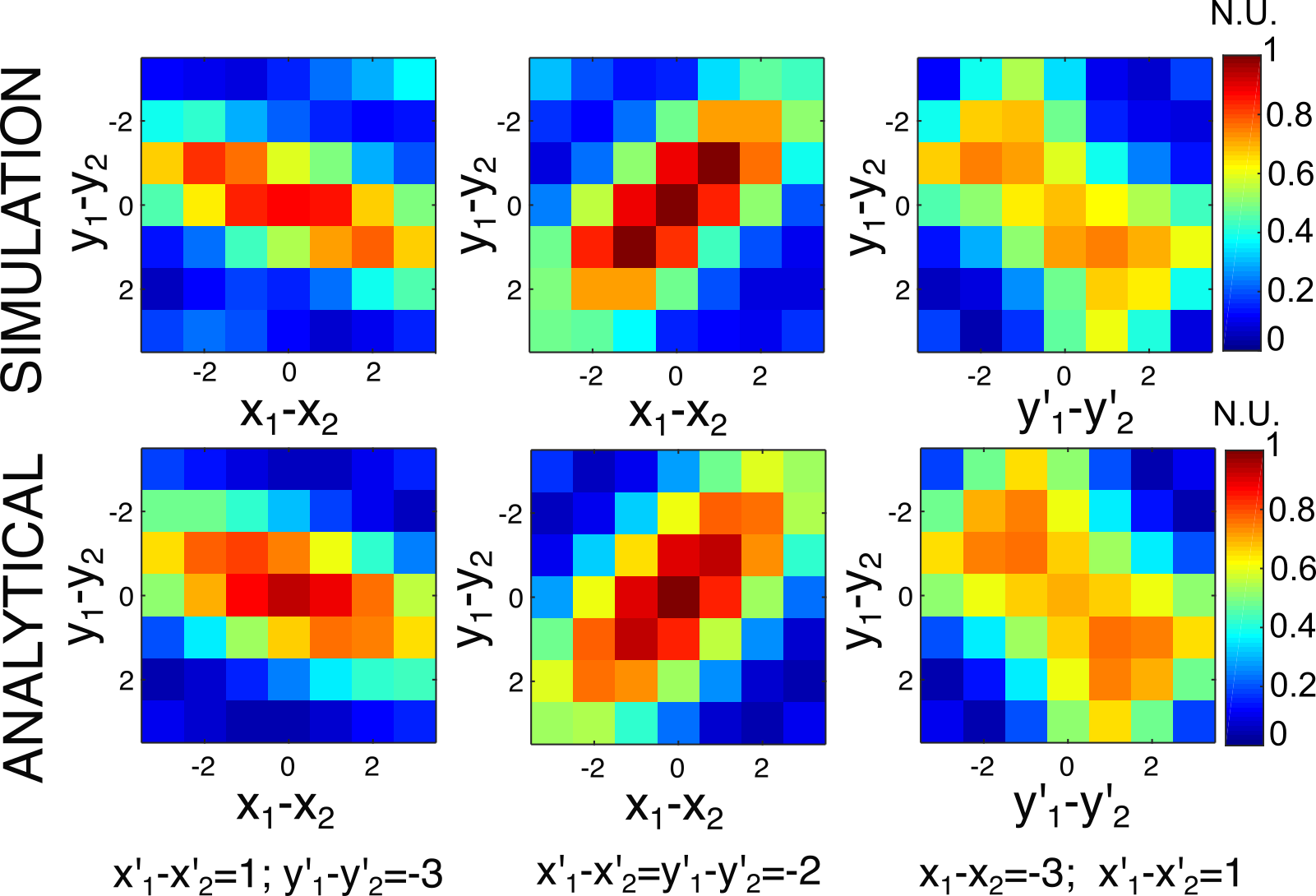}
\end{center}
  \caption{Comparison of analytical predictions and numerical simulations for 4 photon correlations in quantum images. To allow the comparison, the correlations of each row are normalized to unity and represented with a false color scale expressed in normalized units (N.U.).  The coordinates that are not fixed in the figure are averaged out.}
  \label{fig:numerics}
\end{figure}

In Fig. \ref{fig:numerics} we compare the numerical simulations with the analytical predictions of Eq. (\ref{Eq:P4}). 
 The numerical simulations  were carried out for a pump beam with a waist  $w_0=200$ $\mu$m, a crystal thickness of 50 $\mu$m, and a pump laser wavelength of $351$ nm. The pixel size (used to discretize the numerical simulations) is $2.7$ $\mu$m. The field of view is  $256*256$ pixels.
 The intensity of the pump beam is adjusted so that the intensity at the center of the signal and idler beams after propagation through the crystal is approximately $0.6$ photon per pixel. The beam is then propagated $100$ $\mu$m beyond the crystal in order to defocus it. The stochastic simulations were repeated $5\ 10^5$ times in order to obtain sufficient statistics.
 To exhibit the oscillations of $P^{(4)}$  encoded in the variable $S$
 we fix two coordinate differences, use two other coordinate differences as plot variables, and average over the remaining 4 coordinates. 
The differences between theory and numerics correspond to a signal-to-noise ratio of $4.9$, in agreement with a model of the numerical uncertainties developed in Appendix \ref{Appendix:SNR}.

\section{Experimental implementation.}

Experimental parameters for an experimental implementation  could be as follows. $100$ fs pump pulses at $355$ nm, with a waist $w_0=200$ $\mu$m, produce
 photon pairs  in a $\beta$-barium borate crystal of thickness 50$\mu$m. The non-colinear signal and idler photons pass through a $3$ nm wide notch fillter to ensure that they are indistinguishable. They are imaged onto the CCD cameras with  either a 2f or 4f optical system, slightly defocused as discussed in the main text. The overall detection efficiency (the probability that a signal or idler photon is registered on the camera, taking into account all losses and detector inefficiency) can be taken to be  $\eta=0.3$. In Appendix \ref{Appendix:SNR} we estimate that $10^5$ to  $10^6$ images  need to be taken in order to reproduce experimentally images similar to those in 
 Fig. \ref{fig:numerics}.
For comparison Ref. \cite{Defienne18} used $10^7$ images to analyse in detail the biphoton wavefunction in a quantum imaging experiment, showing that such an experiment is accessible using present technology). Two  related experiments are Ref. \cite{Bornman19} that demonstrated  quantum imaging experiments with 4 photons but with images were restricted to  4 pixels, and  \cite{Eibl03} that studied 4 photon correlations but using the polarisation degree of freedom.

\section{Interpretation as entanglement swapping}

The high dimensional space in which the proposed $4$ photon experiment takes place  makes the experiment much richer. As illustration of the new features that  emerge 
we show that a modification of the experiment allows for an interpretation as entanglement swapping \cite{Bennett93}. 
 
Recall that initially the  photon pairs shared between Alice and Bob are entangled, but there is no entanglement between Bob's photons. The joint detection of Alice's two photons projects Bob's photons into an entangled state. The intuition is that the overlap of the wavefunctions of Alice's two photons, followed by the detection of these two photons at specific positions, is analoguous to the
action of the beam splitter followed by joint detection in the teleportation experiment in \cite{Zeilinger97}. 
Indeed, suppose we postselect that two photon pairs are produced and  that Alice's photons are detected at positions ${\bf  x}_{A1}$
and ${\bf  x}_{A2}$. Then it follows from Eq. (\ref{Eq:GaussianPsi}))
that  Bob's two photons are projected onto the entangled state
\begin{eqnarray}
\vert \phi\rangle &=& \int  d {\bf  x}'_1  d {\bf  x}'_2 
\left(
\Phi({\bf  x}_{A1}, {\bf  x}'_1) \Phi({\bf  x}_{A2}, {\bf  x}'_2) \right.
\nonumber\\
& &
\left.\quad \quad 
+ \Phi({\bf  x}_{A1}, {\bf  x}'_2) \Phi({\bf  x}_{A2}, {\bf  x}'_1) \right)
a_{{\bf  x}'_1}^\dagger 
a_{{\bf  x}'_2}^\dagger 
\vert 0 \rangle\ .
\label{eq:phiBB}
\end{eqnarray}

To illustrate this in more detail, suppose that ${\bf  x}_{A1}=(+a,0)$ and 
${\bf  x}_{A2}=(-a,0)$,  that the biphoton wavefunction is given by Eq. (\ref{Eq:Phixx'zz'alphabeta}), and  that  Bob's photons are postselected to be  in the vicinity of $(+l,0)$ and $(-l,0)$. For large enough defocusing, and small enough values of $a$ and $l$, the
quantum state Eq. (\ref{eq:phiBB}) is approximately given by a momentum entangled state (see Appendix \ref{App:Telep} for the derivation):
\begin{equation}
\vert \phi\rangle \approx 
 \left( 
e^{i \varphi_1} 
\vert p'_{-+} ; +l  \rangle
\vert p'_{+-} ; -l \rangle
+
e^{i \varphi_2} 
\vert p'_{--} ; +l  \rangle
\vert p'_{++} ; -l \rangle
\right)
 \label{eq:phiBB2}
\end{equation}
where $\varphi_{1,2}$ are unimportant phases, and 
$\vert   p' ; \pm l   \rangle$ are approximate momentum states located near $(\pm l,0)$  respectively with zero momentum in the $y$ direction, and  momentum 
$ p'_{\pm \pm} = \pm\frac{ \beta l}{2} \pm \frac{ \beta a}{2} $ in the $x$ direction.

\begin{figure}
\begin{center}
  \includegraphics[width=1.0\linewidth]{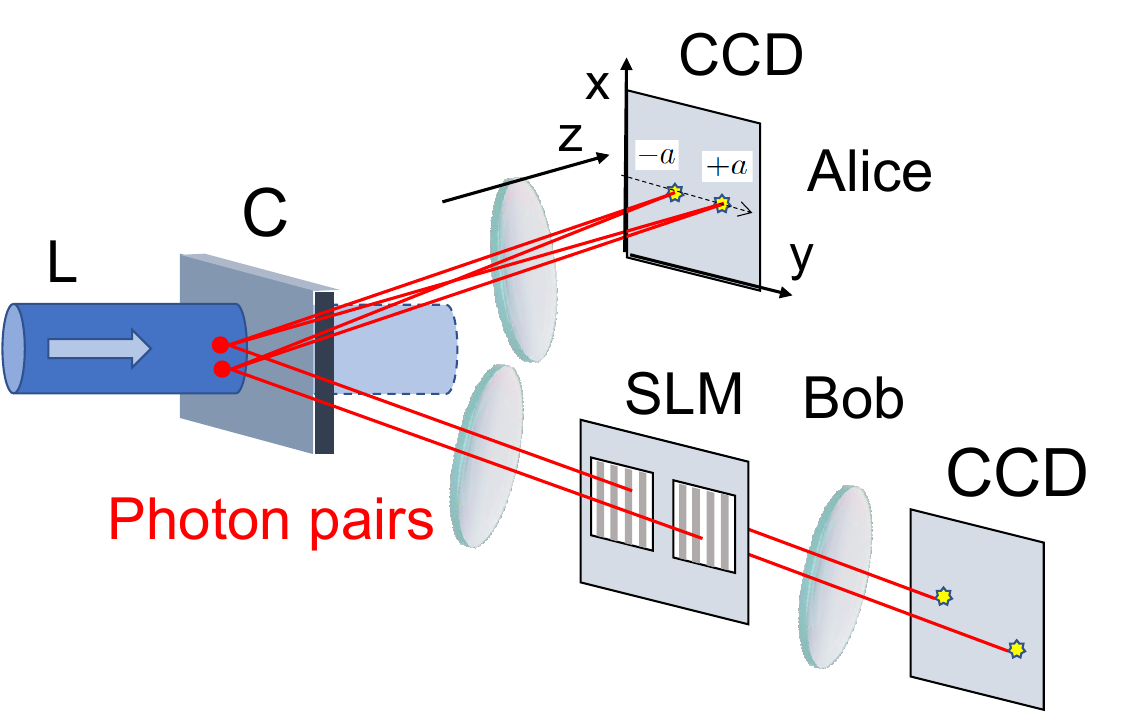}
\end{center}
  \caption{Proposed setup to demonstrate entanglement swapping between quantum images (the lens indicate  schematically that imaging optics is required).
Alice's camera plane is defocused with respect to the imaging the crystal surface ($z$ coordinate in the image).  
  The joint detection of Alice's photons at a specific position, say $(+a,0)$ and $(-a,0)$,  leaves Bob's photons in an entangled state. In order to analyse the entanglement between his photons, Bob uses a Spatial Light Modulator (SLM) on which is imprinted a periodic phase profile (represented by gray lines in the figure).
  The SLM is positioned so as to be in the plane imaging the crystal surface.
   If one photon passes through one window of his SLM, and the other photon passes through the other window, then the periodic phase imprinted on the photons allows Bob to analyse the momentum entanglement between his photons. 
This is obtained by positioning Bob's camera in the far field (where detection events measure the transverse momentum of the photons).
     (The detailed procedure is described in Appendix \ref{App:DemEntSwap}).
}
  \label{fig:teleport}
\end{figure}

In order to  demonstrate that the resulting state indeed has the form Eq. (\ref{eq:phiBB2}), one needs to measure the first photon (located near $(+l,0)$)  in the basis spanned by $\vert p'_{-\pm};l\rangle$   and the second photon 
 (located near $(-l,0)$)  in the basis spanned by $\vert p'_{+\pm};-l\rangle$. Such measurements can be realised by inserting 
 along the paths of the photons a Spatial Light Modulator (SLM) such that around regions $(\pm l,0)$ the SLM has phase profiles which are periodic with period 
$2 \pi / \Delta p$, with $\Delta p = p_{\pm +} - p_{\pm -} = \beta a$, and then measuring in the far field, see Fig. \ref{fig:teleport} and Appendix \ref{App:DemEntSwap}.

\section{Conclusion}

In the present work we have shown that entangled photons of high dimension exhibit interesting multiphoton correlations, focusing on the specific case of  entangled quantum images. 
Interesting multiphoton correlations already appear in the $4$ photon case. They can be exibited by defocusing the images, which is of course  easy experimentally. We provide detailed analytical predictions for the resulting $4$ photon intereferences. These are supported by numerical simulations.
We further show that this experiment can be interpreted as entanglement swapping between photons. Bob's photons are initially unentangled. But the joint detection of Alice's photons projects Bob's photons onto an entangled state.

The present work could be extended in several directions. First of all it calls for an experimental demonstration as 
the  multiphoton correlations described here are accessible with current experimental techniques. The main experimental difficulties are to make the different photons produced indistinguishible (except for the spatial degree of freedom) in order to allow for multiphoton intereferences, and to accumulate sufficient statistics in order to see the correlations emerge from the background.
As discussed above, this seems of comparable difficulty to other experiments that have been realised previously.
A successful experiment of $4$-photon correlations would set the stage for investigating higher order correlations (6 and more photons).

Second, the proposed experiment should be compared with boson sampling experiments\cite{Broome13,Spring13,Tillmann13,Paesani19,Zhong20} whose aim is to have a highly complex bi-photon wavefunction 
in order to maximise the complexity of a classical simulations. Here we propose using a quite simple bi-photon wavefunctions, leading to simple expressions for the multi-photon correlations. However the bi-photon wavefunction can be complexified, for instance by inserting SLMs along the optical path as proposed in \cite{Raoul21}. Since quantum imaging experiments with hundreds to thousands of modes have been demonstrated\cite{Edgar12,Moreau14}, quantum imaging may ultimately provide a  more scalable approach to Boson sampling.

Third, the interpretation as entanglement swapping suggests that many mutliphoton experiments such as generation of GHZ states \cite{Bouwmeester99}, W states\cite{Eibl04}, etc...   could find analog implementations using quantum imaging experiments.

Finally, the general approach proposed here is not limited to the spatial degrees of freedom. Photons entangled in other degrees of freedom, such as frequency or angular momentum, could also be used to investigate multiphoton correlations.
These  directions in which the present work can be extended show that multiphoton correlations betweeen quantum images promises to be a rich area of study, both theoretically and experimentally.

\appendix

\section{Robustness}\label{AppendixRobustness}

In this section we show that in the limit of large defocusing our results do not depend on the gaussian approximation used in the main text. We derive a form for the biphoton wavefunction and for the 4 photon coincidence probabilities that is valid when $z+z'$ is large. 

In order to compare the results obtained in this section with the results obtained in the main text, we note that when  $z+z'$ is large Eq. (\ref{Eq:Phixx'zz'alphabeta}) in the main text takes the form
\begin{eqnarray}
& & \Phi({\bf  x}, {\bf  x'};z,z') \nonumber\\
&\approx &  \exp\left( - \frac{k^2}{b^2(z+z')^2} \vert {\bf  x} - {\bf  x'}\vert^2 + i \frac{k}{2(z+z')} \vert {\bf  x} - {\bf  x'}\vert^2 \right)\nonumber\\
\ .
\label{Eq:Phiapprox}
\end{eqnarray}

In general the biphoton wavefunction at transverse positions ${\bf  x}$ and  ${\bf  x'}$ on Alice and Bob's camera's, which are positioned a distance $z$ and $z'$ from the non linear crystal, is given by
\begin{eqnarray}
&\Phi({\bf  x}, {\bf  x'};z,z') 
\propto \int\int d {\bf  p}  d {\bf  p'}  \ 
\tilde \Phi(  {\bf  p}  , {\bf  p'}  )& \nonumber\\
& \times
\exp\left( 
- i \frac{z}{2k}\vert {\bf  p} \vert^2 - i \frac{z'}{2k}\vert {\bf  p'} \vert^2 
+ i {\bf  p}   \cdot {\bf  x} + i {\bf  p'}   \cdot {\bf  x'} 
\right) &
\label{Eq:PhiApp-1}
\end{eqnarray}
where $\Phi(  {\bf  p}  , {\bf  p'}  )$ is the biphoton wavefunction (in momentum space) at the crystal surface, which we no longer take to be gaussian.

We place ourselves in the translation invariant limit, so that
\begin{equation} 
\tilde \Phi(  {\bf  p}  , {\bf  p'}  )
= \delta (  {\bf  p}  +  {\bf  p'}  ) \tilde f(  \frac{ {\bf  p}  - {\bf  p'}  }{2})\ .
\label{Eq:PhiApp-2}
\end{equation}
Eq. (\ref{Eq:PhiApp-1}) therefore becomes
\begin{eqnarray}
&\Phi({\bf  x}, {\bf  x'};z,z') 
\propto \int d {\bf  p}   \ 
\tilde f(  {\bf  p}  )& \nonumber\\
&\times 
\exp\left( 
- i \frac{(z+z')}{2k}\vert {\bf  p} \vert^2 
+ i {\bf  p}   \cdot ( {\bf  x}   -{\bf  x'} )
\right)& 
\label{Eq:PhiApp-3}
\end{eqnarray}
We suppose that the crystal is not very thick, so that at the crystal surface the photons are highly correlated in position. This implies that  $f$ (the Fourier transform of $\tilde f$) is strongly peaked around $0$, and hence that  $ \tilde f$ is a slowly varying function.
Therefore, for large enough $z+z'$, the integral in Eq. (\ref{Eq:PhiApp-3}) can be approximated by saddle point integration. The saddle is at
\begin{equation}
{\bf  p} ^* = \frac{ k ( {\bf  x}   -{\bf  x'} )}{z+z'}
\label{Eq:PhiApp-4}
\end{equation}
and $\Phi({\bf  x}, {\bf  x'};z,z') $ is approximately given by
\begin{equation}
\Phi({\bf  x}, {\bf  x'};z,z') 
\approx 
\tilde f\left( \frac{ k ( {\bf  x}   -{\bf  x'} )}{z+z'}\right)
\exp\left( 
 i \frac{k}{2(z+z')}\vert {{\bf  x}   -{\bf  x'} } \vert^2 
\right) \ .
\label{Eq:PhiApp-5}
\end{equation}
This can be compared with Eq. (\ref{Eq:Phiapprox}).
We see that the quadractic phase of $\Phi$ is robust prediction of the model. On the other hand the gaussian prefactor is not.

Upon inserting Eq. (\ref{Eq:PhiApp-5}) into Eq. (\ref{Eq:4photon}),  one finds that the 4 photon correlation probability takes the  form
\begin{eqnarray}
& & P^{(2)}({\bf  x}_1,{\bf  x}_2; {\bf  x'}_1,{\bf  x'}_2; z,z') \nonumber\\
&\propto&
\vert \tilde f_{11'} \vert^2 \vert \tilde f_{22'} \vert^2 
+\vert \tilde f_{12'} \vert^2 \vert \tilde f_{21'} \vert^2 
\nonumber\\
& &+ 2 \vert \tilde f_{11'} \tilde  f_{22'}  \tilde f_{12'} \tilde  f_{21'} \vert
\cos \left( 
  \frac{k}{z+z'}
  ({\bf  x_1}   -{\bf  x}'_1 )\cdot ({\bf  x_2}   -{\bf  x}'_2 ) + \varphi
\right) \nonumber\\
\label{Eq:PhiApp-6}
\end{eqnarray}
where we use the notation
\begin{equation}
\tilde f_{ij'} = \tilde f\left( \frac{ k ( {\bf  x}_i   -{\bf  x}'_j )}{z+z'}\right) 
\end{equation}
for the slowly varying prefactors, and $\varphi$ is the phase of
$\tilde f_{11'} \tilde f_{22'}  \tilde f_{12'}^* \tilde f_{21'}^* $.

Equation (\ref{Eq:PhiApp-6}) has the same structure as Eq. (\ref{Eq:P4}) in the main text. In particular the last term in Eq.  (\ref{Eq:PhiApp-6})  corresponds to the oscillating term $ \cos(\beta S/4)$  in Eq. (\ref{Eq:P4}). The argument of the harmonic function is the same ($\beta S/4$) in both expressions, up to the phase  $\varphi$.

Further note that when ${\bf  x}_1, {\bf  x}_2, {\bf  x}'_1, {\bf  x}'_2$ are all close to each other, then we have approximate equality  of  the prefactors $\tilde f_{11'} = \tilde f_{22'} =  \tilde f_{12'} = \tilde f_{21'} = \tilde f({\bf  0})$, consequently $\varphi = 0$, and therefore Eq. (\ref{Eq:PhiApp-6}) further simplifies to 
\begin{eqnarray}
& & P^{(2)}({\bf  x}_1,{\bf  x}_2; {\bf  x'}_1,{\bf  x'}_2; z,z') \nonumber\\
 &\propto& 2 \vert \tilde f({\bf  0})\vert^4 \left ( 1 + 
\cos \left( 
  \frac{k}{z+z'}
  ({\bf  x_1}   -{\bf  x}'_1 )\cdot ({\bf  x_2}   -{\bf  x}'_2 ) \right)
\right) \ .\nonumber\\
\label{Eq:PhiApp-6B}
\end{eqnarray}

Eqs. (\ref{Eq:PhiApp-6}) and (\ref{Eq:PhiApp-6B}) show that
the oscillations in the 4 photon probabilities is thus a robust prediction of the proposed experiment.

\section{Higher order correlations}\label{App:HighOrder}

In this section we give expressions for  higher order correlations in quantum imaging experiments, i.e. between $n$ photons on Alice's camera and $n$ photons on Bob's camera. The case $n=2$ yields Eq. (\ref{Eq:P4}) in the main text.

The amplitude to find $n$ photons on Alice's camera at positions 
${\bf  x}_1,\cdots,{\bf  x}_n$ and $n$ photons on Bob's camera at positions 
${\bf  x'}_1,\cdots,{\bf  x'}_n$ is given by 
\begin{eqnarray}
& &\langle 0 \vert a_{{\bf  x}_1} \cdots a_{{\bf  x}_n} a_{ {\bf  x'}_1}\cdots 
a_{{\bf  x'}_n} \vert \Psi \rangle \nonumber\\
&\propto &  {\mathrm{Perm}}\left(
\Phi_{{\bf  x}_1,\cdots,{\bf  x}_n; {\bf  x'}_1,\cdots,{\bf  x'}_n} \right) 
\nonumber\\
& = &
\sum_\sigma \exp \left( -\frac{\alpha - i \beta}{4} \sum_{i=1}^n \vert x_i - x'_{\sigma(i)}\vert^2\right) 
\nonumber\\
&=&\exp \left( -\frac{\alpha - i \beta}{4}   D^{(n)} \right)  \sum_\sigma 
\exp \left ( -  \frac{\alpha - i \beta}{4}  S^{(n)}_\sigma \right)\nonumber\\
\label{Eq:Amp(n)}
\end{eqnarray} 
where we have used Eq. (\ref{Eq:Phixx'zz'alphabeta}) in the main text for the biphoton wavefunction, and where
\begin{eqnarray}
D^{(n)}  &=& \frac{1}{n}\sum_{i,j=1}^n \vert x_i - x'_{j}\vert^2
\label{Supp:1}\\
S^{(n)}_\sigma &=&  \sum_{i=1}^n  \left(  \vert x_i - x'_{\sigma(i)}\vert^2 - \frac{1}{n}\sum_{j=1}^n \vert x_i - x'_{j}\vert^2\right) \nonumber\\
&=&
2 \sum_{i=1}^n \left( x_i \cdot x'_{\sigma(i)} - \frac{1}{n}\sum_{j=1}^n  x_i \cdot x'_{j} \right)\ .
\label{Supp:2}
\end{eqnarray}
(Note that $D^{(2)}=2D$ where $D^{(2^)}$ is defined in Eq. (\ref{Supp:1}) and $D$ is defined in the main text below Eq. (\ref{Eq:P4})).

We therefore find that 
\begin{widetext}
\begin{eqnarray}
 P^{(n)}({\bf  x}_1,\cdots,{\bf  x}_n; {\bf  x'}_1,\cdots,{\bf  x'}_n)
&\propto & \exp( -   \frac{\alpha}{2} D^{(n)} )
\vert \sum_\sigma \exp\left (-  \frac{\alpha - i \beta}{4} S^{(n)}_\sigma \right) \vert^2\nonumber\\
&=&\exp( -  \frac{\alpha}{2} D^{(n)} )
\left( \sum_\sigma  \exp\left (-  \frac{\alpha}{2}  S^{(n)}_\sigma \right)
\right.\nonumber\\
& & \left.
+ 2 \sum_{\sigma< \sigma'}
\exp\left (-  \frac{\alpha}{4}  ( S^{(n)}_\sigma  + S^{(n)}_{\sigma'}) \right)
\cos \left ( 
\frac{\beta}{4}  ( S^{(n)}_\sigma -  S^{(n)}_{\sigma'}) \right)
\right)\nonumber\\
\label{Eq:P(n)}
\end{eqnarray}
\end{widetext}
(where by $\sum_{\sigma< \sigma'}$ we mean that we do a double sum over all permutations, with $\sigma \neq \sigma'$, and each pair $(\sigma, \sigma')$ only counted once).

 Eq. (\ref{Eq:P(n)}) contains $\frac{n! (n!+1)}{2}$ terms.  In the case $n=2$ the expression simplifies because there are only two permutations, the identity $I$ and $\sigma_{12}$, and also because we have that $S^{(2)}_I = - S^{(2)}_{\sigma_{12}} $. This  yields  Eq. (\ref{Eq:P4}) in the main text. In the case $n=3$ there are 6 permutations, corresponing to 21 terms. The complexity of the multiphoton correlations thus grows rapidly as the number of photons increases.

\section{Signal-to-noise ratio (SNR)}\label{Appendix:SNR}

Here we estimate the Signal to Noise Ratio (SNR) in the proposed experiment, and hence the number of camera frames required to reach a desired SNR. We present a qualitative estimate that shows the dependence on the main parameters. For instance
our estimates are only valid for low or moderate number of produced pairs, and we do not take into account effects due to the interference effects described in the main text (this is precisely the signal we want the measure). A more precise estimate would compute exactly all the probabilities for the signal we want to measure and all the backgrounds. This goes beyond the present work.

\subsection{Parameters}

For ease of reading, we list here the parameters that will be used in our analysis.

\begin{itemize}
\item $\overline N$ is the average number of photons produced by the pump pulse.

\item $n_{Pixels} \gg 1$ is the number of pixels over which photons can be registered. Note that this does not necessarily correspond to the number of pixels of the camera as one may bin several camera pixels together, and on the other hand part of the camera area may not be used.

\item $0< \eta < 1$ is the probability that a photon is detected. ($1 -\eta$ are the losses, including all optical losses, detector efficiency, etc..).

\item
$n_{PixelsCond}>1 $ is the number of pixels  over which Bob's photon can be found, given that Alice detected a photon at  a specific pixel.

\item $P_1$ denotes the probability that a photon is registered on  a pixel of Alice's camera (or a pixel  of Bob's camera).  

\item $P_2^{Coincid}$  denotes the probability that two photon from a pair are registered, one  on Alice's camera and one on Bob's cameras.

\item $P_4^{Coincid}$  denotes the probability that four photon from two pairs are registered, two  on  Alice's camera and two on  Bob's cameras.

\item $n^{frames}$ is the number of camera frames accumulated to get sufficient statistics.

\item $P_{dark}$ is the probability of a dark count. We will take $P_{dark}=0$  below (supposing that it is not the dominant source of noise). We indicate below how to take into account $P_{dark}\neq 0$.

\item $n^{temp}$ is the number of temporal/spectral modes of Alice and Bob's photons.  We will initially suppose that there is a single temporal mode (i.e that the pump pulse is sufficiently short, and subsequent spectral filtering of signal and idler sufficiently narrow, that the downconverted photons cannot be distinguished based on temporal-spectral information). We will then show how our estimates change when there is more than one temporal mode.

\end{itemize}

\subsection{Single pixel detection probability}

The probability of having a click on a given pixel $i$ of Alice's camera (or $i'$ of Bob's camera) is
\begin{equation}
 P_1(i)=P_1(i' )= \eta   \frac{\overline  N }{n_{Pixels}} \ .
\end{equation}

\subsection{Two photon coincidence probability}

If Alice registers a photon at pixel $i$, then the other photon of the pair can be registered over a certain zone $Z_i$ of the Bob's camera. Let us consider the probability of a coincidence $(i,i')$ where $i'$ belongs to the zone $Z_i$. 
\begin{eqnarray}
P_2^{Coincid}(i,i')&=&P_1(i)P^{Coincid} (i'\vert i)\nonumber\\
&=& P_1(i) \frac{\eta}{n_{PixelsCond}}
\end{eqnarray}
where 
 $P^{Coincid} (i'\vert i)$ is the probability that Bob detects the partner photon at pixel $i'$, given that Alice detected a photon from the same pair at pixel $i$.

\subsection{Four photon coincidence probability}

What is the probability that Alice detects photons at pixels $i,j$ and Bob detects photons at pixels $i',j'$? The interesting case is when $i',j'\in Z_i$, that is photon $i'$  could be the partner of photon $i$ or of $j$, and similarly for $j'$. Then there can be interferences between the different pairs. This is  given by 
\begin{equation}
P_4^{Coincid}(ij,i'j')=P_2^{Coincid}(i,i')P_2^{Coincid}(jj')
\end{equation}
(up to order 1 factors due to the interferences described in the main text, which is precisely what we want to measure).
Thus 
\begin{equation}
P_4^{Coincid}= \left( P_2^{Coincid}\right)^2 =\left(  P_1 \frac{\eta}{n_{PixelsCond}} \right)^2\ .
\end{equation}



\subsection{Total coincidence probabilities}

The probability of a click on one pixel is
\begin{equation}
P(click) = 
P_1 + P_{dark}
\end{equation}
where we indicate how to take into account the dark counts. We neglect $P_{dark}$ in what follows, but it could be easily be included in the estimates of the SNR.

The total probability of a coincidence $(i,i')$ where $i'$ belongs to the zone $Z_i$ on one image is
 \begin{equation}
P^{CoincidTotal}_2(i,i') =  P^{Coincid}_2(i,i') +
\left( P(click) \right)^2 
\label{Eq:Pcoin2}
\end{equation}
where the second term is due to accidental coincidences.

The probability of a 4-fold coincidence is 
\begin{eqnarray}
& & P^{CoincidTotal}_4(ij,i'j'\vert \mathrm{1\ image}) \nonumber\\ &=& 
P^{Coincid}_4(ij,i'j'\vert \mathrm{1\ temp\ mode})
 +
\left( P(click) \right)^4 +...\nonumber\\
\label{Eq:Pcoin4}
\end{eqnarray}
where the second term is due to accidental 4 fold coincidences (There are other accidental kinds of 4 fold coincidences, for instance when 2 photons belong to a pair, and the other 2 do not, for simplicity we do not write all these terms).

\subsection{SNR for 2 Photon correlations}

To measure the correlations, we need to accumulate $n^{frames}$ camera images.

The number of single detections on pixel $i$ is
 \begin{eqnarray}
 N_1(i) &=& n^{frames} P(click) \pm \sqrt{n^{frames} P(click) }
 \end{eqnarray}
where we add the statistical uncertainty.

The number of coincidences on pixels $i,i'$ follows from Eq.  (\ref{Eq:Pcoin2}):
 \begin{eqnarray}
 N_2^{coincid}(i,i') &=& 
n^{frames} P^{Coincid}_2  + n^{frames}  P(click)^2\ .\nonumber\\
\label{Eq:Ncoin2}
 \end{eqnarray}
 The signal we want to measure is
\begin{equation}
\text{S}_2 = n^{frames} P^{Coincid}_2
\end{equation} 
while the noise is the statistical fluctuations of the two terms in Eq. (\ref{Eq:Ncoin2}):
\begin{equation}
\text{N}_2 = \pm \sqrt{ n^{frames}  P^{Coincid}_2} +
\pm \sqrt{n^{frames}  P(click)^2}\ .
\end{equation} 
The first noise term will dominate when we have low pump power so that photon pairs are rare, while the second noise term will dominate when photon pairs are common. The two noise terms are of comparable magnitude when
$P(click)^2 = P^{Coincid}_2$ which corresponds to 
\begin{equation}
\overline{N} = \frac{n_{Pixels}}{n_{PixelsCond}} \ .
\label{Eq:OptimalN}
\end{equation}
That is the two noise terms are comparable when approximately one pair is produced per zone of size $n_{PixelsCond}$.
Since the first noise term is the fluctuations of the signal, to improve the SNR ratio we should increase the pump power (i.e. increase $\overline{N}$) until the second noise term becomes comparable to the first. From now on we assume that this is the case, and that the first noise term is smaller or equal than the second.

The Signal to Noise Ratio is then
\begin{eqnarray}
\text{SNR}_2
&=&  \sqrt{n^{frames} }  \frac{ P^{Coincid}_2 }
{  P_1} \ .
\label{Eq:SNR2}
\end{eqnarray}

Hence the number of frames needed to exhibit 2 photon coincidences is
\begin{eqnarray}
n^{frames}_2  &=& \text{SNR}_2^2 \frac{  P_1}{ P^{Coincid}_2  } 
\nonumber\\
&=&  \text{SNR}_2^2  \frac{n_{PixelsCond}^2 }{ \eta^2}\ .
\label{Eq:nframes2}
\end{eqnarray}

\subsection{SNR for 4 Photon correlations}

Similarly the number of 4 fold coincidences on pixels $i,j,i',j'$ follows from Eq.  (\ref{Eq:Pcoin4}):
\begin{eqnarray}
  & & N_4^{Coincid}(ij,i'j') 
 \nonumber\\
  &=& n^{frames} P_4^{Coincid}(ij,i'j') 
+n^{frames}  P(click)^4 +...\ .\nonumber\\
\label{Eq:Ncoin4}
 \end{eqnarray}

 The signal we want to measure is
\begin{equation}
\text{S}_4 = n^{frames} P^{Coincid}_4
\end{equation} 
while the noise is the statistical fluctuations of the two terms in Eq. (\ref{Eq:Ncoin4}):
\begin{equation}
\text{N}_4 = \pm \sqrt{ n^{frames}  P^{Coincid}_4} +
\pm \sqrt{n^{frames}  P(click)^4}\ .
\end{equation} 

One easily shows that two noise terms are comparable when Eq. (\ref{Eq:OptimalN}) is satisfied. To make the SNR ratio maximal, one should work in a regime where the pump powe is large enough that the first noise term is smaller or equal than the second.

The Signal to Noise Ratio is then
\begin{eqnarray}
 \text{SNR}_4 &=&
\sqrt{n^{frames} } \frac{ P_4^{Coincid} } {  P_1^2 }  
\nonumber\\
&=&\sqrt{n^{frames} }  \left( \frac{P_2^{Coincid }} {  P_1 }\right)^2 
\nonumber\\
&=&\sqrt{n^{frames} }  \frac{ \eta^2} {  n_{PixelsCond}^2 }
\end{eqnarray}

Hence the number of frames needed to exhibit 4 photon coincidences is
\begin{eqnarray}
n^{frames}_4  &=& \text{SNR}_4^2  \left( \frac{  P_1}{ P_2^{Coincid }} \right)^4
\nonumber\\
&=&  \text{SNR}_4^2  \frac{n_{PixelsCond} ^4}{ \eta^4}
\label{Eq:nframes4}
\end{eqnarray}

\subsection{Effect of distinguishable photons}

If one uses a long pump pulse (or equivalently a too broad specral filter), then photon pairs produced at different times will be distinguishable. 
This situation would also arise if the pump pulses were short enough, but the camera averages over several successive pump pulses.
We denote $n_{temp}$ the number of temporal modes that are average over in one camera frame.

Then we have that that $P_1$, $P_2^{Coincid }$, $P_4^{Coincid} $ are all mutiplied by 
$n_{temp}$.

Therefore the factor $n_{temp}$ cancels in the SNR for photon pairs Eq. (\ref{Eq:SNR2}). Hence one can  study the photon pair correlations using a CW pump (which is often done experimentally).

However the factor  $n_{temp}$  does not cancel in the SNR for 4 photon coincidences. Indeed only a fraction $1 / n_{temp}$ 4 fold coincidences will come from indistinguishable pairs while all other 4 fold coincidences will come from distinguishable pairs and will contribute to background but not to the desired signal. Hence we will have
\begin{equation}
 \text{SNR}_4 \to \frac{ \text{SNR}_4}{\sqrt{n_{temp}}}
\end{equation}
and
\begin{equation}
 n^{frames}_4  \to n_{temp} n^{frames}_4 \ .
\end{equation}

\subsection{Estimation of the number of frames needed}\label{SubSec:NFrames}

We assume the following parameters:
\begin{eqnarray}
\eta &=& 0.3\nonumber\\
n_{PixelsCond} &=& 30 \nonumber\\
\text{SNR} &=& 10\nonumber\\
\frac{n_{Pixels}}{n_{PixelsCond}} &=&10^3
\end{eqnarray}
The last estimate expresses the fact that the total size of a camera image is much larger than the zone over which photons are correlated. Hence a single camera image contains effectively 
$\frac{n_{Pixels}}{n_{PixelsCond}}$ independent images, each covering a zone of size $n_{PixelsCond}$, and the number of frames that need to be taken is reduced by this factor.

Hence from Eq. (\ref{Eq:nframes2} )we have 
\begin{equation}
n^{frames}_2 = 10^3
\end{equation}
and from Eq. (\ref{Eq:nframes4})
\begin{equation}
n^{frames}_4 = 10^{7}\ .
\end{equation}

This estimate is reduced if one wants to obtain a figure such as Fig.  \ref{fig:numerics} in the main text, as in this figure $K\approx 50$ four-fold correlations are averaged to obtain each pixel in the figure. To obtain the same SNR,  the number of frames required is reduced by a factor $K$ (see discussion in subsection \ref{SubSec:Comp}). We thus reach an estimate between $10^5$ and $10^6$ frames to reproduce experimentally a figure such as Fig.  \ref{fig:numerics}.

\subsection{Comparison between analytics and  simulations in Fig. \ref{fig:numerics} of the main text.}\label{SubSec:Comp}

In the simulation, we perform for each pixel of Fig. \ref{fig:numerics} an average of $K=49$ values of four-fold correlations between Alice's pixels of coordinates $\bf{x_1,x_2}$, and Bob's pixels of coordinates $\bf{x'_1,x'_2}$. The averaged values correspond to an unique value of 
$\bf{x_1-x_2, x'_1-x'_2}$. This averaging multiplies the SNR by $\sqrt{S}$, giving, in the conditions of the simulation corresponding to $n_{temp}=1$, $\eta=1$, an expected SNR of:
\begin{equation}
SNR=\frac{\sqrt{S}\sqrt{n_{frames}}}{n_{pixelsCond}^2}
\label{SNRsimull}
\end{equation}
With $n_{frames}=5.10^5$, $n_{pixelsCond}=32$, we obtain with Eq.\ref{SNRsimull} an SNR of $4.8$. This value of $n_{pixelsCond}$ seems reasonable because of the decreasing of the correlations close to the edges of the sub-figures in figure 2.  

We can compare this with the SNR estimated from the images by comparing the 
 analytical and the simulated values in Fig. \ref{fig:numerics}. This gives an estimated SNR of:
\begin{equation}
SNR_{est}=\frac{\overline{Analytic}}{\sqrt{\overline{(Analytic-simulated)^2}}}=4.9
\end{equation}
There is thus a good agreement betwen the two estimates.

\section{Entanglement Swapping}\label{App:Telep}

We give here details about the interpretation in terms of entanglement swapping.

We suppose that the biphoton wavefunction is given by Eq. (\ref{Eq:Phixx'zz'alphabeta})  and that Alice detects her photons at positions ${\bf  x}_{A1}=(+a,0)$ and 
${\bf  x}_{A2}=(-a,0)$. We denote the transverse coordinates on Bob's detection plane by
${\bf  x}'_1 = (x'_1,y'_1)$ and ${\bf  x}'_2 = (x'_2,y'_2)$.
Then the wavefunction of Bob's two photons takes the form (insert Eq. (\ref{Eq:Phixx'zz'alphabeta}) into Eq. (\ref{eq:phiBB})):
\begin{widetext}
\begin{eqnarray}
 \vert \phi \rangle  &=&
\exp\left( -\frac{\alpha - i \beta}{4} \left( \left( a - x'_1\right)^2 + {y'_1}^{2}
+\left( a + x'_2\right)^2 + {y'_2}^{2}\right)
\right.
+ \left.
\exp\left( -\frac{\alpha - i \beta}{4} \left( \left( a + x'_1\right)^2 + {y'_1}^{2}
+\left( a - x'_2\right)^2 + {y'_2}^{2}\right)\right)
\right)
\nonumber
\\
& = &
\exp \left(  -\frac{\alpha - i \beta}{4} \left(  {y'_1}^{2} + {y'_2}^{2}\right) \right)
\exp \left(  -\frac{\alpha - i \beta}{4} \left( 2 a^2 +  {x'_1}^{2} + {x'_2}^{2}\right) \right)
\nonumber\\ & & \times 
\left( 
\exp\left(
-\frac{\alpha - i \beta}{2}   a (x'_2 - x'_1) \right)
+
\exp\left(
-\frac{\alpha - i \beta}{2}   a (x'_1 - x'_2) \right)
\right)
\nonumber\\
\label{Eq:phiBob}
\end{eqnarray}
\end{widetext}

We suppose that ${\bf  x}'_1= (x'_1,y'_1)$
and ${\bf  x}'_2= (x'_2,y'_2)$ are located in the vicinity of $(l,0)$ and $(-l,0)$. We
then write
\begin{equation}
x'_1=l+\delta'_1 \quad \quad x'_2=-l+\delta'_2 \ .
\end{equation}
We bound the region in which Bob's particles can be located by
\begin{eqnarray}
&-\delta \leq \delta'_1 , \delta'_2 \leq + \delta& \nonumber\\
& -y \leq y'_1 , y'_2 \leq +y 
\end{eqnarray}
with \begin{equation}
\delta \ll  l\ .
\end{equation}

In order to simplify Eq. (\ref{Eq:phiBob}) we make the following assumptions
\begin{eqnarray}
\beta &\gg & \alpha 
\quad \text{(large defocusing, i.e. $Z \gg 1$)} \nonumber\\
(\alpha , \beta ) \times y &\ll& 1 
\quad \text{(neglect $y$ dependence)} \nonumber\\
(a^2, al, l^2)\times \alpha &\ll& 1 \quad \text{(neglect all terms that depend on $\alpha$)} \nonumber\\
\delta ^2  \beta &\ll& 1 \quad \text{(neglect second order terms in  $\delta'_1, \delta'_2$)} \nonumber\\
( a^2, al, l^2)\times \beta &\gg& 1 
\quad \text{(keep phases proportional to $\beta$)\ .} 
\nonumber
\end{eqnarray}
With these assumptions Eq. (\ref{Eq:phiBob}) takes the form
\begin{eqnarray}
 \phi  
& \simeq  &
\exp \left( i\frac{ \beta}{2} \left(  a^2 + l^2\right) \right)
\nonumber\\
& \times& 
 \left( 
\exp\left( i \frac{ \beta}{2} \left( -2a l + (l-a) \delta'_1 - (l-a) \delta'_2 \right)\right)
\right.
\nonumber\\
& & \left. +
\exp\left( i \frac{ \beta}{2} \left( +2a l + (l+a)  \delta'_1 - (l+a) \delta'_2 \right)\right)
\right)\nonumber\\
\label{Eq:phiBobCC}
\end{eqnarray}
which we can rewrite in terms of momentum states as
\begin{widetext}
 \begin{eqnarray}
& \vert \phi \rangle  
& \simeq  
\exp \left( i\frac{ \beta}{2} \left(  a^2 + l^2\right) \right)
 \left( 
\exp\left( -i \beta a l \right)
\vert p'_1 = -\frac{ \beta l}{2} + \frac{ \beta a}{2} ; +l\rangle
\vert p'_2 = +\frac{ \beta l}{2} - \frac{ \beta a}{2} ; -l\rangle
\right.
\nonumber\\ & & \quad \quad 
\left. +
\exp\left( +i \beta a l \right)
\vert p'_1 = -\frac{ \beta l}{2} - \frac{ \beta a}{2} ; +l\rangle
\vert p'_2 = +\frac{ \beta l}{2} + \frac{ \beta a}{2} ; -l\rangle
\right)\ .
\label{Eq:phiBobDD}
\end{eqnarray}
\end{widetext}
This is the expression given in the main text in Eq. (\ref{eq:phiBB2}).

\section{Demonstrating  Entanglement Swapping}\label{App:DemEntSwap}

In the main text we skteched how to demonstrate experimentally that state Eq. (\ref{Eq:phiBobDD}) is entangled by inserting a SLM in the beam of Bob's photons, see Fig. \ref{fig:teleport} in the main text. We present here the argument in more detail.

The state Eq. (\ref{Eq:phiBobDD})  is a two qubit state, which we can write in abstract notation as
\begin{equation}
 \vert \phi \rangle  = a\vert 0\rangle_B\vert 0\rangle_{B'} + b\vert 1\rangle_B\vert 1\rangle_{B'}\ .
 \label{Eq:EntAbst}
\end{equation}
where 
\begin{eqnarray}
a &=& \exp\left( -i \beta a l \right) \nonumber\\
b &=& \exp\left( +i \beta a l \right) \nonumber\\
\vert 0\rangle_B &=& \vert p'_1 = -\frac{ \beta l}{2} + \frac{ \beta a}{2} ; +l\rangle \nonumber\\
\vert 1\rangle_B &=& \vert p'_1 = -\frac{ \beta l}{2} - \frac{ \beta a}{2} ; +l\rangle\nonumber\\
\vert 0\rangle_{B'} &=&\vert p'_2 = +\frac{ \beta l}{2} - \frac{ \beta a}{2} ; -l\rangle\nonumber\\
\vert 1\rangle_{B'} &=& \vert p'_2 = +\frac{ \beta l}{2} + \frac{ \beta a}{2};-l \rangle\ .
\label{Eq:EntAbstParam}
\end{eqnarray}
and where the subscripts $B$ and $B'$ denote the photons that are located near $+l$ and $-l$ respectively.

Measuring in the $\{\vert 0\rangle_B, \vert 1\rangle_B\}$ and $\{\vert 0\rangle_{B'}, \vert 1\rangle_{B'}\}$ bases is straightforward. First insert a mirror to separate spatially the $B$ states from the $B'$ states (recall that these states are localised in momentum and in space). Then 
 put the CCD camera in the far field.

But measuring only in the computational basis (the basis $\{\vert 0\rangle, \vert 1\rangle\}$) is not enough to demonstrate entanglement. For this we need additional measurements. We show how to do so using a Spatial Light Modulator (SLM).

Suppose that we put on the SLM a periodic phase profile $\varphi(x)=\epsilon \cos (k x + \theta)$. Then a wavefunction $\psi(x)$ becomes
\begin{equation}
\psi(x) \to \psi(x) e^{i \varphi(x)}\ .
\end{equation}
We can expand the phase in Fourier series as
\begin{eqnarray}
e^{i \epsilon \cos (k x + \theta)} &=& \sum_n a_n(\epsilon)  e^{i n (k x+ \theta)}\nonumber\\
&\approx& 1 + i \frac{\epsilon}{2} e^{i (k x + \theta)} + i \frac{\epsilon}{2}  e^{-i (k x + \theta)} 
+ O(\epsilon^2)
\label{Eq:JacobiAnger}
\end{eqnarray}
where the exact coefficients $a_n(\epsilon) = i^n J_n(\epsilon)$ follow from the Jacobi-Anger expansion (with $J_n$ the Bessel function of the first kind).
 In the second line we give the expression for small $\epsilon$ which we use below as it is conceptually simpler, and sufficient to demonstrate the principle.

From Eq. (\ref{Eq:JacobiAnger}) it follows that acting on a momentum state $\vert p \rangle$, the SLM carries out the transformation
\begin{equation}
\vert p \rangle \to \vert p \rangle + i \frac{\epsilon e^{i \theta}}{2} \vert p - k \rangle  + i \frac{\epsilon e^{-i \theta}}{2} \vert p + k \rangle
\end{equation}

Acting on the superposition of two momentum states $\vert \psi \rangle = \alpha \vert p \rangle +
\beta \vert p + k \rangle$ (where we suppose that the momenta differ by exactly the wave number $k$ of the  SLM phase), we therefore have
\begin{eqnarray}\vert \psi \rangle =
\alpha \vert p \rangle +
\beta \vert p + k \rangle
&\to &
 i \frac{\epsilon e^{i \theta}}{2}  \alpha \vert p -k\rangle 
\nonumber\\ & & 
 +
\left(   \alpha +  i \frac{\epsilon e^{i \theta}}{2}   \beta \right) \vert p  \rangle
\nonumber\\ & & +
\left(   \beta +  i \frac{\epsilon e^{-i \theta}}{2}   \alpha \right)  \vert p +k\rangle
\nonumber\\ & & + 
 i \frac{\epsilon e^{-i \theta}}{2}  \beta \vert p + 2k \rangle
 \label{Eq:momentumA}
\end{eqnarray}
By measuring in the far field, the probability of finding the photon in spots corresponding to momenta $p$ and $p+k$ will be equal to the norm square of the coefficients of the second and third line in Eq. (\ref{Eq:momentumA}).
These probabilities are proportional to
\begin{eqnarray}
&\vert \left( \langle p \vert  -  i \frac{\epsilon e^{-i \theta}}{2}  \langle p +k \vert 
\right) \vert \psi \rangle \vert^2& \nonumber\\
&\quad \texttt{and} \quad& \nonumber\\
&
\vert \left(   -  i \frac{\epsilon e^{+i \theta}}{2}  \langle p  \vert  + \langle p + k \vert
\right) \vert \psi \rangle \vert^2&
\label{Eq:Probas-1}
\end{eqnarray}
respectively.

Therfore by both measuring in the original $\{\vert p \rangle, \vert p +k\rangle\}$ basis, and by carrying out the above measurement for different values of 
$\theta$ (for fixed $\epsilon$) one easily obtains tomographically complete information on the state. 

Note that using the  Jacobi-Anger expansion mentioned in Eq. (\ref{Eq:JacobiAnger}), one can carry out the above analysis for finite value of $\epsilon$. One finds for instance that the probabilities for the far field probabilities at momenta $p$ and $p+k$ are given by
\begin{eqnarray}
&\vert \left( a_0^* \langle p \vert  + a_{1}^* e^{-i \theta}  \langle p +k \vert 
\right) \vert \psi \rangle \vert^2& \nonumber\\
&\quad \texttt{and} \quad& \nonumber\\
&\vert \left(   a_{-1}^*  e^{+i \theta}\langle p  \vert  + a_0^* \langle p + k \vert
\right) \vert \psi \rangle \vert^2&
\label{Eq:Probas-2}
\end{eqnarray}
instead of the approximate expression Eq. (\ref{Eq:Probas-1}).

Going back to the problem Eqs. (\ref{Eq:EntAbst}, \ref{Eq:EntAbstParam}), we see that choosing $k= \beta a$ for the wave number of the phase on the SLM will allow to do a tomographically complete set of measurements on the state 
 Eq. (\ref{Eq:EntAbst}).

\end{document}